\documentstyle[psfig,11pt]{article}
\def\pn{\par\noindent}

\def\gsimeq
{\hbox{\raise0.5ex\hbox{$>\lower1.06ex\hbox{$\kern-1.07em{\sim}$}$}}}
\def\lsimeq
{\hbox{\raise0.5ex\hbox{$<\lower1.06ex\hbox{$\kern-1.07em{\sim}$}$}}}
\def\pn{\par\noindent}

\begin{document}
\vspace{1.0cm}

{\Large \bf A QUALITATIVE TEST OF A UNIFIED MODEL OF SEYFERT GALAXIES WITH $BEPPOSAX$}

\vspace{1cm}

M. Cappi$^{1*}$, L. Bassani$^{1}$, G. Malaguti$^{1}$, G.G.C. Palumbo$^{1,2}$, 
M. Dadina$^{3}$, A. Comastri$^{4}$, G. Di Cocco$^{1}$, P.Blanco$^{5}$, 
D. Dal Fiume$^{1}$, A. Fabian$^{6}$, F. Frontera$^{1,7}$, M. Guainazzi$^{8}$, 
T. Maccacaro$^{9}$, R. Maiolino$^{10}$, G. Matt$^{11}$, L. Piro$^{12}$, 
M. Trifoglio$^{1}$, and N. Zhang$^{13}$

\vspace{1.0cm}
$^{1}$ Istituto Te.S.R.E./CNR, Via Gobetti 101,I-40129 Bologna, Italy\\
$^{2}$ Dipartimento di Astronomia, Universita` di Bologna, Italy\\
$^{3}$ BeppoSAX S.D.C., ASI, Via Corcolle 19, I-00131 Rome, 
Italy\\
$^{4}$ Osservatorio di Bologna, Bologna, Italy\\
$^{5}$ UCSD, San Diego Ca, USA\\
$^{6}$ Institute of Astronomy, Cambridge University, Madingley Road,
Cambridge CB3 0HA,UK \\
$^{7}$ Dipartimento di Fisica, Universita' di Ferrara, Via Paradiso 11, I-
44100 Ferrara, Italy\\
$^{8}$ Astrophysics Division, SCD -ESA, ESTEC, Postbus 299, 2200 AG
Noordwijk, The Netherlands\\
$^{9}$ Osservatorio Astronomico di Brera, Via Brera 28, I20121 Milano, Italy\\
$^{10}$ Osservatorio Astrofisico di Arcetri, Via L.E. Fermi 5, I5015 
Firenze, Italy\\
$^{11}$ Dipartimento di Fisica,Universita' di RomaIII, Via della Vasca Navale 
84, I00146 Roma, Italy\\
$^{12}$ Istituto di Astrofisica Spaziale, Via Del Fosso del Cavaliere,
I-001333 Roma, Italy \\
$^{13}$ University Space Research Association, Huntsville-Al, USA\\
\hspace{1truecm}{$^*$ e-mail: mcappi@tesre.bo.cnr.it}

\vspace{0.5cm}

\section*{ABSTRACT}

The broad band 0.1-200 keV spectra of a sample of 5 Seyfert 2 
galaxies (NGC 7172, Mkn 3, NGC 2110, NGC 4507 and NGC 7674) 
have been measured within the first year of the $BeppoSAX$ Core program.
All sources have been detected up to $\sim$ 100 keV 
and their spectral characteristics derived with good accuracy.
Although the results obtained from the detailed analysis of individual 
sources indicate some ``source-by-source'' differences, we show 
in the following that all spectra are consistent, at least qualitatively, 
with what expected from a ``0$^{th}$-order'' version of unified models. 
Indeed, a simple test on these data 
indicates that these Seyfert 2 galaxies are on average intrinsically 
very similar to Seyfert 1 galaxies (i.e., steep at E$\gsimeq$ 10 keV) 
and that the main difference can be 
ascribed to a different amount of absorbing matter along the line of 
sight (i.e. different inclinations of our line of sight with respect to a 
putative molecular torus or different thicknesses of the tori).

\section{\bf PREDICTIONS FROM UNIFIED MODELS}

The discovery of broad emission lines in the polarized optical
spectra of several Seyfert 2 galaxies (Antonucci \& Miller 1985, 
Tran et al. 1992) has provided the basis for
a unified model of Seyfert galaxies in which the main discriminating
parameter between Seyfert 1 and Seyfert 2 nuclei is the inclination with respect to
our line of sight of a supposed obscuring torus surrounding the
central source (see Antonucci 1993 for a review).
In this scheme, Seyfert 1 galaxies are active nuclei observed nearly
perpendicularly to the torus plane (unabsorbed) whereas Seyfert 2 galaxies
represent those seen through the torus (absorbed) and, of course, the 
main prediction is that Seyfert 2 galaxies are {\it intrinsically} similar 
to Seyfert 1 galaxies, once the effects of the torus are properly accounted for.

It is now widely recognized that the {\it intrinsic} high energy spectrum of 
Seyfert 1 galaxies consists, on average, of a steep power-law continuum 
($\Gamma$ $\sim$ 1.9--2.0, Nandra \& Pounds 1994) with an exponential cut-off 
typically at energies larger than 150 keV (Zdziarski et al. 1996, 
Perola et al. 1998). Therefore, one would expect that Seyfert 2 galaxies 
exhibit similar high-energy properties.
X-ray observations (mainly below $\sim$ 20 keV) 
of Seyfert 2 galaxies, have 
shown a variety of spectral characteristics not always consistent with 
a ``0$^{th}$-order'' version of unified models (Smith \& Done 1996, Cappi et al. 
1996, Turner et al. 1998). However, at high energies where the effects of 
absorption and matter reprocessing are less evident, measurements are sparse for 
Seyfert 2 galaxies (but see Johnson et al. 1997). It has been therefore difficult 
to assess the Seyfert 1 nature of the primary spectrum of Seyfert 2 galaxies 
from these observations.

Based on these general considerations, we have undertaken a program of 
observations with $BeppoSAX$, aimed at studying the X-ray spectral properties 
of bright Seyfert 2 galaxies over a broad energy band (up to about 200 keV) and at 
testing the validity of unified models. 
As a matter of fact, $BeppoSAX$ can provide crucial information on the 
intrinsic source properties because high energy photons from a few 
to several keV can pass through the circumnuclear intervening material and can 
therefore be compared to the ``typical'' spectrum of Seyfert 1 galaxies.
So far 5 objects have been observed within the first AO cycle, namely
NGC 7674 (Malaguti et al. 1998a), Mkn 3 (Cappi et al.1998), NGC 2110 
(Malaguti et al. 1998b), NGC 7172 and NGC 4507 (see also Bassani et al. 1998).

\section{A SIMPLIFIED QUALITATIVE TEST}

A simple, qualitative, test has been performed on our sources by fitting 
each source of the sample with the same model: a soft power-law component 
plus a hard, absorbed, power-law component with reflection and associated 
Fe K line as illustrated in Fig. 1. In the framework of unified models, 
the soft power-law is attributed to scattered soft X-ray emission from 
ionized material placed above the molecular torus, while 
the hard X-ray and reflection components are interpreted as the direct 
component absorbed by the torus and 
its reflection from the inner side of the torus, respectively.
The intensity of the soft, scattered, component was assumed to be 
$\sim$ 2\% that of the direct one. The intensity of the 
reflection component was fixed to R = 1 (conrresponding to a 2$\pi$ 
coverage as viewed from the X-ray source), and the iron line 
was assumed to be produced through both the reflection and absorption 
with an equivalent width of $\sim$ 1 keV (with respect to the 
reflected continuum) and $\sim$ 500 $\times$ $N_{\rm H}\over{1.23 \times 10^{24}}$ eV
(with respect to the direct, absorbed, component), as expected from theoretical 
models (George \& Fabian 1991, Leahy \& Creighton 1993).
The only free parameters were the intensity and photon index 
of the direct continuum and the absorption column along the line of sight.

The fit results and unfolded spectra obtained from this test are 
given in Figure 2. The most interesting result is that all sources 
are well described by a steep, Seyfert 1-like spectrum, with 
$\Gamma$ $\sim$ 1.79--1.95.
This is a newly discovered behaviour of Seyfert 2 galaxies for energies 
up to $\sim$ 100 keV which supports unified models. It is stressed that 
this result is largely independent on the presence of the steep soft component and 
on the assumed intensity of the reflection component. 
This is demonstrated by the fact that 
the average Seyfert 2 spectrum obtained averaging all the PDS 20-200 keV data 
(except NGC 7674) can be well fitted ($\chi^2$ $\sim$ 1.2) by 
a single power-law model with $\Gamma_{20--200 keV}$ = 1.85 $\pm$ 0.05
and shows no deviation from the power-law up to $\sim$ 150 keV.
The observed major differences in the quality of fits ($\chi^{2}_{red}$ ranging 
from 0.9--1.8) are to be ascribed to two main factors. The first is source 
by source differences in the Fe K complex which indicates the need 
of a more dedicated analysis. The second depends upon the different 
amount of absorbing matter along the line of sight.
The less absorbed source is NGC 2110 which also shows less indication 
for a reflection component and the most absorbed one is NGC 7674, where only 
the reflection component is observed possibly because the direct component 
is completely blocked by a Compton thick molecular torus 
(with $N_{\rm H} \gsimeq 10^{25}$ cm$^{-2}$, Malaguti et al. 1998a). 
Intermediate cases are NGC 7172, NGC 4507 and Mkn 3; in the latter, 
both the direct and reflected components are clearly resolved spectroscopically.
Moreover although a detailed measurement in individual sources of the 
high-energy cutoff is difficult, it appears clear in the data (see Figure 2) 
that there is no evidence of such a cutoff for energies up to at least 
$\sim$ 100--150 keV.

{\bf REFERENCES}
\vspace{-5mm}
\begin{itemize}
\setlength{\itemindent}{-8mm}
\setlength{\itemsep}{-1mm}

\item []
Antonucci, R.R.J., 1993, $ARA\&A$, {\bf 31}, 473
\item []
Antonucci, R.R.J. \& Miller, J.S., 1985, {\it ApJ}, {\bf 297}, 621
\item []
Bassani, L., et al. 1998, to appear in proceedings of ``Dal nano- al tera
-eV: tutti i colori degli AGN'', third Italian conference on AGNs, Roma, 
{\it Memorie S.A.It}, astro-ph/9809327
\item []
Cappi M., Mihara, T., Matsuoka, et al., 1996, {\it ApJ}, {\bf 456}, 141
\item []
Cappi, M., et al. 1998, $A\&A$, in press, astro-ph/9902022
\item []
George, I.M. \& Fabian, A.C., 1991, {\it MNRAS}, {\bf 249}, 352
\item []
Johnson, W.N., Zdziarski, A.A., Madejski, G.M., Paciesas, W.S., Steinle, H., \& Lin Y-C, 
1997, in proceedings of the Fourth Compton Symposium, Ed. D. Dermere, M.S. Stcikman and 
J.D. Kurfess, {\it AIP}, 283
\item []
Leahy D.A. and Creighton J., 1993, {\it MNRAS}, {\bf 263}, 314
\item []
Malaguti, G., et al. 1998a, $A\&A$, {\bf 331}, 519
\item []
Malaguti  G. et al. 1998b, $A\&A$, in press, astro-ph/9901141
\item []
Nandra, K. \& Pounds, K.A., 1994, {\it MNRAS}, {\bf 268}, 405
\item []
Perola, C., et al. 1998,  to appear in proceedings of ``Dal nano- al tera
-eV: tutti i colori degli AGN'', third Italian conference on AGNs, Roma, {\it Memorie S.A.It}
\item []
Smith, D.A., \& Done, C., 1996, {\it MNRAS}, {\bf 280}, 355
\item []
Tran, H.D., Miller, J.S. \& Kay, L.E., 1992, {\it ApJ}, {\bf 397}, 452
\item []
Turner, T.J., George, I.M., Nandra, K., \& Mushotzky, R.F., 1998, {\it ApJ}, {\bf 493}, 91
\item []
Zdziarski, A.A., Johnson, W.N., Poutanen, J., Magdziarz, P., \& Gierlinski, M., 
1996, in ``The Transparent Universe'', proceedings of the 2nd INTEGRAL Workshop, 
Ed. C. Winkler, T.J.-L. Courvoiser and P. Durouchoux, ESA SP-382, 373

\end{itemize}

\begin{figure}[htb]
\psfig{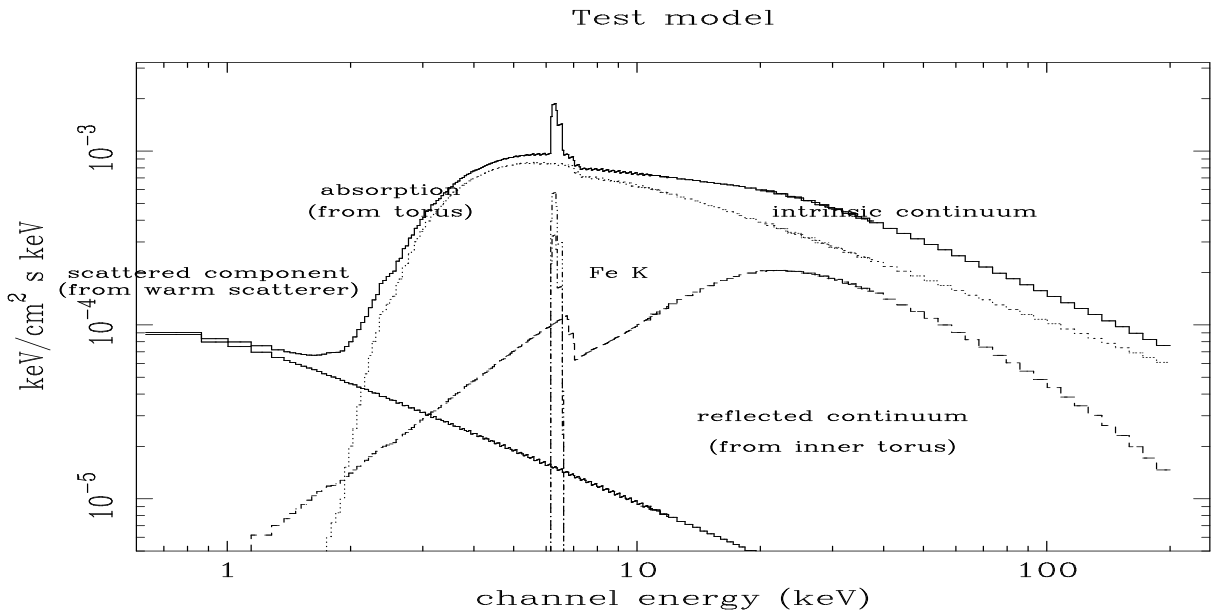}
\caption{Simple, qualitative, test model used to fit all the sources of the 
sample in the framework of unified models.}
\end{figure}
\pn

\vfill\eject


\normalsize
\begin{figure}[htb]
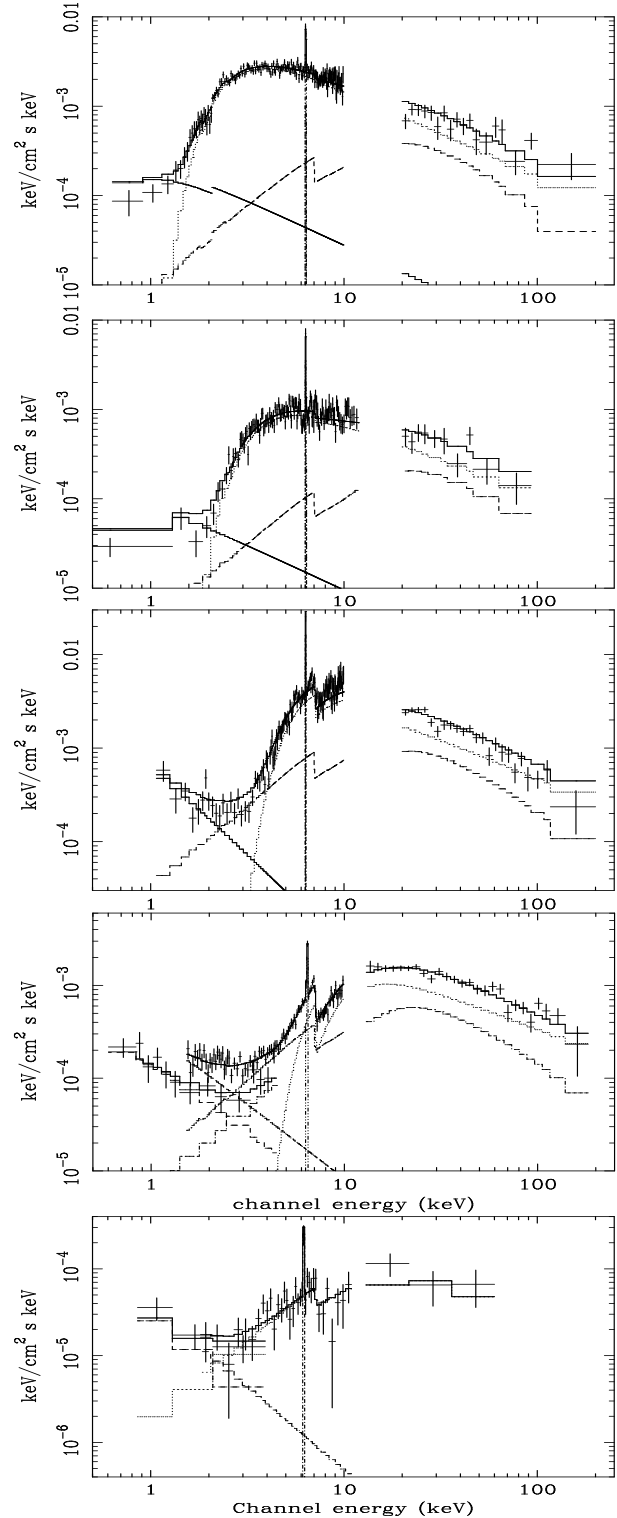

\hspace{1.5truecm}
\parbox{5truecm}
{{\bf NGC 2110}:
\pn
$N_{\rm H}$ $\sim$ 5 $\times$ 10$^{22}$ cm$^{-2}$
\pn
$\Gamma$ $\sim$ 1.97 \hspace{2.5truecm}{$\Longleftrightarrow$}
\pn
$\chi^{2}_{red}$ $\sim$ 1.16
\pn
Malaguti et al. 1998, submitted to A\&A}
\  \hspace{0truecm}     \
\parbox{8truecm}
{\psfig{file=./cappi_fig2.ps,width=8cm,height=4cm,angle=-90}}
\pn
\hspace{1.5truecm}
\parbox{5truecm}
{{\bf NGC 7172}:
\pn
$N_{\rm H}$ $\sim$ 10$^{23}$ cm$^{-2}$
\pn
$\Gamma$ $\sim$ 1.83 \hspace{2.5truecm}{$\Longleftrightarrow$}
\pn
$\chi^{2}_{red}$ $\sim$ 1.4
\pn
Dadina et al., in prep}
\  \hspace{0truecm}     \
\parbox{8truecm}
{\psfig{file=./cappi_fig3.ps,width=8cm,height=4cm,angle=-90}}
\pn
\hspace{1.5truecm}
\parbox{5truecm}
{{\bf NGC4507}:
\pn
$N_{\rm H}$ $\sim$ 5 $\times$ 10$^{23}$ cm$^{-2}$
\pn
$\Gamma$ $\sim$ 1.83 \hspace{2.5truecm}{$\Longleftrightarrow$}
\pn
$\chi^{2}_{red}$ $\sim$ 1.8 
\pn
Bassani et al., in prep}
\  \hspace{0truecm}     \
\parbox{8truecm}
{\psfig{file=./cappi_fig4.ps,width=8cm,height=4cm,angle=-90}}
\pn
\hspace{1.5truecm}
\parbox{5truecm}
{{\bf Mkn 3}:
\pn
$N_{\rm H}$ $\sim$ 1.3 $\times$ 10$^{24}$ cm$^{-2}$
\pn
$\Gamma$ $\sim$ 1.79 \hspace{2.5truecm}{$\Longleftrightarrow$}
\pn
$\chi^{2}_{red}$ $\sim$ 0.9
\pn
Cappi et al., submitted to A\&A}
\  \hspace{0truecm}     \
\parbox{8truecm}
{\psfig{file=./cappi_fig5.ps,width=8cm,height=4cm,angle=-90}}
\pn
\hspace{1.5truecm}
\parbox{5truecm}
{{\bf NGC 7674}:
\pn
$N_{\rm H}$ $\gsimeq$ 10$^{25}$ cm$^{-2}$
\pn
$\Gamma$ $\sim$ 1.95 \hspace{2.5truecm}{$\Longleftrightarrow$}
\pn
$\chi^{2}_{red}$ $\sim$ 0.9
\pn
Malaguti et al. 1998a}
\  \hspace{0truecm}     \
\parbox{8truecm}
{\psfig{file=./cappi_fig6.ps,width=8cm,height=4cm,angle=-90}}
\caption{Fit results and broad-band unfolded spectra obtained from the qualitative test.
See text for a description of the model fitted. $N_{\rm H}$ increases going from the 
top (NGC 2110) to the bottom (NGC 7674).}
\end{figure}

\end{document}